\def\ts     {\thinspace}
\def\msol     {\ifmmode{{\rm M}_{\odot}}\else{M$_{\odot}$}\fi}
\def\lsol     {\ifmmode{{\rm L}_{\odot}}\else{L$_{\odot}$}\fi}
\def\zsol     {\ifmmode{{\rm Z}_{\odot}}\else{Z$_{\odot}$}\fi}
\def\hh     {\ifmmode{{\rm H}_2}\else{H$_2$}\fi}
\def\ha     {\ifmmode{{\rm H}\alpha}\else{${\rm H}\alpha$}\fi}
\def\nh   {\ifmmode{N(\hi)}\else{$N$(\hi)}\fi}
\def\kms    {\ifmmode{{\rm \ts km\ts s}^{-1}}\else{\ts km\ts s$^{-1}$}\fi}
\documentstyle[12pt,aasms4]{article}

\newcommand{\ea}{\emph{et al.\ }}
\newcommand{\etal}{\emph{et al.\ }}
\newcommand{\hi}{\ion{H}{1}}
\newcommand{\msun}{M$_\odot$}

\slugcomment{revised version submitted to the AJ}

\lefthead{Walter et al.}
\righthead{Atomic and Molecular Gas in NGC~3077}

\begin{document}
\title{The Interacting Dwarf Galaxy NGC\,3077: \\
The Interplay of Atomic and Molecular Gas with Violent Star Formation}

\author{Fabian Walter} 
\affil{Owens Valley Radio Observatory, California Institute of Technology, 
Pasadena, CA, 91125}
%\\Electronic mail: fw@astro.caltech.edu}

\author{Axel Wei\ss}
\affil{Radioastronomisches Institut der
Universit\"{a}t, Auf dem H\"{u}gel 71, D--53121 Bonn, Germany}
%\\Electronic mail: aweiss@astro.uni-bonn.de} 

\author{Crystal Martin
\footnote{Visiting astronomer Kitt Peak National Observatory}}
\affil{Astronomy Department, California Institute of Technology, 
Pasadena, CA, 91125}
%\\Electronic mail: clm@astro.caltech.edu}

\and 
\author{Nick Scoville}
\affil{Astronomy Department, California Institute of Technology, 
Pasadena, CA, 91125}
%\\Electronic mail: nzs@astro.caltech.edu

\begin{abstract}
We present a comprehensive multi--wavelength study of the nearby
interacting dwarf galaxy NGC\,3077 (member of the M\,81 triplet). High
resolution VLA \ion{H}{1}\ observations show that most of the atomic
gas ($\sim$90\%) around NGC\,3077 is situated in a prominent tidal arm
with a complex velocity structure. Little \ion{H}{1}
($\sim5\times10^7$\msun) is associated with NGC\,3077 itself.  High
resolution OVRO observations of the molecular component (CO) reveal
the presence of 16 molecular complexes near the center of NGC\,3077
(total mass: $\sim1.6\times10^6$\msun). A virial mass analysis of the
individual complexes yields a lower CO--to--H$_2$ conversion factor in
NGC\,3077 than the Galactic value - a surprising result for a dwarf
galaxy. The lower conversion factor can be explained by extreme
excitation conditions and the metallicity of the molecular gas. The
total (atomic and molecular) gas content in the centre of NGC\,3077 is
displaced from the stellar component of NGC\,3077 -- this implies that
not only the gas at large galactocentric radii is affected by the
interaction within the triplet but also the center. We speculate that
the starburst activity of NGC\,3077 was triggered by this
redistribution of gas in the center: H$\alpha$ as well as Pa$\alpha$
images show the presence of violent central star formation as well as
dramatic ionized supershells reaching galactocentric distances of
$\sim$1\,kpc. Some of these supershells are surrounded by neutral
hydrogen. In a few cases, the rims of the ionized supershells are
associated with dust absorption.
%Surprisingly, star formation,
%coinciding spatially as well as kinematically with diffuse molecular
%gas, is also found 500\,pc west of the central star forming region. 
The most prominent star forming region in NGC\,3077 as probed by
Pa$\alpha$ observations is hidden behind a dust cloud which is traced
by the molecular complexes. Correcting for extinction we derive a star
forming rate of 0.05 M$_\odot$\,year$^{-1}$, i.e. given the reservoir
in atomic and molecular gas in NGC\,3077, star formation may proceed
at a similar rate for a few 10$^8$ years. The efficiency to form stars
out of molecular gas in NGC\,3077 is similar to that in M\,82.

\end{abstract}


\section{Introduction}
\label{Intro}

NGC\,3077 is a prominent example of an interacting dwarf
galaxy. Together with the prototypical starburst galaxy M\,82 and the
spiral galaxy M\,81 it forms the famous M\,81 triplet. Although an
interaction between the three galaxies is not at all obvious at
optical wavelengths (but see Arp 1965, Getov \& Georgiev 1988) it has
been long known through observations in the
\ion{H}{1} line that the 3 galaxies are interacting: \hi\ tidal arms
and tails connect the three galaxies over a projected distance of more
than 70 kpc (Cottrell 1977, van der Hulst 1979, Yun \etal\ 1994). The
outstanding star forming activity in M\,82 is usually attributed to
the gravitational interaction with M\,81 and/or NGC\,3077.

Although the molecular and starburst properties of M\,82 have been the
subject of innumerable studies, little is known about its starforming
neighbor, NGC\,3077. This is surprising since it has been long known
that NGC\,3077 is experiencing intense star formation (e.g., Barbieri
\etal 1974). As noted by Price \& Gullixson (1989) it is difficult to
classify NGC\,3077 on basis of its optical morphology. Emission line
studies of NGC\,3077 (e.g., Price \& Gullixson 1989; Thronson, Wilton
\& Ksir 1991 (TWK91); Martin 1997, 1998, 1999) show that streamers and
supershells of ionized gas surround a bright core of H$\alpha$
emission. This core is also associated with soft X--ray emission,
consistent with intense star formation in the center (Bi et al.\
1994). Molecular gas in the center was detected by Becker et al.\
(1989) using the IRAM 30\,m telescope. First interferometric maps of
the molecular component of NGC\,3077 using the OVRO millimeter array
have been presented by Thronson \& Carlstrom (1992) and Meier, Turner
\& Beck (2001). Some general information about NGC\,3077 is compiled in
Tab.~1.

The \ion{H}{1}\ in NGC\,3077 is strongly disrupted by the interaction
with its two neighbors M\,81 and M\,82: most of the neutral gas is
located in a prominent tidal arm east of NGC\,3077 (e.g., van de Hulst
1979, Yun 1994, Walter 1999). Walter \& Heithausen (1999, see also
Heithausen \& Walter 2000) recently discovered a massive molecular
complex in this tidal feature.  Numerical simulations suggest that the
gas in the tidal arm has been stripped off the outskirts of NGC\,3077
during the latest close encounter with M81, some 3 $\,\times\,10^8$ yr
ago (Brouillet \etal\, 1990, Thomasson \& Donner 1993, Yun et al.\
1993). This is in contrast to the suggestion of some authors (Price \&
Gullixson 1989, Bi et al. 1994, TWK91), that NGC\,3077 accreted
(`stole') \ion{H}{1} from M\,81 in this recent interaction. In the
following, we will assume a distance to NGC\,3077 of 3.2\,Mpc (Tammann
\& Sandage 1968). We note that this value has recently refined to be
3.9\,Mpc (Sakai \& Madore 2001), i.e., 20$\%$ higher, however we
decided to use the old distance to make it easier to compare our
results to earlier studies.

In this paper, we attempt to shed more light on the atomic and
molecular gas as well as the starburst properties of NGC\,3077 and
present a multi--wavelength study of this fascinating galaxy. In
Section~2 we describe the observations of the atomic gas (\ion{H}{1},
VLA), molecular gas (CO, OVRO), optical broad and narrow band (KPNO)
and the NIR (NICMOS HST). The different wavelengths are then compared
and discussed in detail in Section~3. Section~4 summarizes the most
important results of our study and relates the gas and star formation
properties of NGC\,3077 to those of its neighbor, M\,82.

\begin{table}
\caption{General information on NGC\,3077}
\label{3077_info}
\begin{center}
%\scriptsize
\begin{tabular}{lc}
\tableline \tableline
Object & NGC\,3077 (UGC\,5398) \nl
Right ascension (J2000.0)\tablenotemark{a} & 10$^{\rm h}$03$^{\rm m}$19.2\fs3 \nl
Declination (J2000.0)\tablenotemark{a} & 68$^{\circ}$ 43$'$ 59$''$\nl
Adopted distance\tablenotemark{b} & 3.2 Mpc \nl
Scale & 15.5 pc/$''$ \nl
Systemic velocity\tablenotemark{c}  & 5 \kms \nl
Apparent magnitude (B)\tablenotemark{d} & 10.6 mag\nl
Distance modulus & m$_{\rm B}$-M$_{\rm B}$=27.6 \nl
Corrected absolute magnitude (B)\tablenotemark{d} & -17.0 mag \nl
Blue luminosity L$_{\rm B}$ & 9.3$\times10^8$ L$_{\rm B\odot}$ \nl
\ion{H}{1}\ mass M$_{\rm HI}$\tablenotemark{c} & $3-10\times10^7$ M$_{\odot}$ \nl \ion{H}{1}\ mass to blue light ratio M$_{\rm HI}$/L$_{\rm B}$ & 0.03--0.1 M$_{\odot}$/L$_{\rm B\odot}$\nl
H$\alpha$ luminosity\tablenotemark{e}&$4.6\times10^{39}$\,erg\,s$^{-1}$\nl
Pa$\alpha$ luminosity\tablenotemark{e}&$9.5\times10^{38}$\,erg\,s$^{-1}$\nl
star formation rate\tablenotemark{e}& 0.05 \msun\,yr$^{-1}$  \nl
H$_2$ mass\tablenotemark{f}& $8\times10^5$ \msun\nl
\tableline
\tableline
\end{tabular}
\tablenotetext{a}{RC\,3, de Vaucouleurs et al.\ 1991}
\tablenotetext{b}{see text}
\tablenotetext{c}{depending on the aperture, see Sec.~3.1}
\tablenotetext{d}{magnitudes are only corrected for extinction within the Milky Way, see TWK91}
\tablenotetext{e}{this study, see Sec.~3.4, note that the H$\alpha$ flux published by Price \& Gullixson (1989) is off by a factor of 20 (TWK91)}
\tablenotetext{f}{this study, see Sec.~3.2.2}
\end{center}
\end{table}

\section{Observations and Data presentation}

\subsection{CO--observations}

We observed NGC\,3077 in the CO(1$\to$0) transition using the Owen's
Valley Radio Observatory's mm array (OVRO) in C, L and H
configurations. In total, 44 hours were spent on source; the
observational details are listed in Table~\ref{ovro_obs}.  Data were
recorded using two simultaneous correlator setups resulting in
velocity resolutions of 5 and 1.3\kms\ (after Hanning smoothing) with
a total bandwidth of 320 and 80 \kms, respectively. Unless otherwise
mentioned, the results presented in this paper were derived using the
1.3\kms\ resolution data. Flux calibration was determined by observing
1328+307 (3C286) and Neptune for approximately 20 minutes during each
observing run.  These calibrators and an additional noise source were
used to derive the complex bandpass corrections. The nearby
calibrators 1031+567 (1.1\,Jy) was used as secondary amplitude and
phase calibrator. The data for each array were edited and calibrated
separately with the {\sc mma} and the {\sc aips} packages. The {\it
uv--}data were inspected and bad data points due to either poor
atmospheric coherences or shadowing were removed, after which the data
were calibrated.

Two sets of datacubes were produced using the task {\sc imagr} in {\sc
aips}, each of them {\sc clean}ed to a level of two times the rms
noise (H\"ogbom 1974, Clark 1980). The first data set was made with
natural weighting, leading to a resolution of $3.7''
\times 3.0''$ ($57\times46$\,pc) and emphasizing large scale
structures (noise: 16\,mJy\,beam$^{-1}$; 0.13\,K in a 5\kms wide
channel, 35\,mJy\,beam$^{-1}$; 0.29\,K in a 1.3\kms\ channel). A second cube
was produced using the {\sc robust} weighting scheme (Briggs 1995).
Eventually, we chose a value of {\sc robust}=0.25, resulting in a
beamsize of $2.4'' \times 1.9''$ ($37\times29$\,pc) and an rms noise
of 22 mJy beam$^{-1}$ (0.44\,K, 5\,\kms wide channels). This latter
resolution starts to resolve individual giant molecular clouds (GMCs)
in NGC\,3077.

To separate real emission from noise when deriving moment maps we
applied the following procedure: the natural weighted data cube was
convolved to a circular beam with a FWHM of $10''$. The smoothed map
was then tested at the $2\sigma$ level; if a pixel fell below this
level, the counterpart in the cube was blanked. After that, the
remaining peaks were inspected. Emission that was present in 3
consecutive channels was considered to be real while all other
remaining spikes were considered to be noise and blanked. The final
result was named the {\sc master} cube. This mask was used to blank
the original {\sc natural} and {\sc robust} data cubes. This method
ensures that the same regions are included when inspecting cubes at
different resolutions and with different signal--to--noise ratios. To
derive physical properties, the CO data were primary beam corrected.

\begin{table}
\caption{OVRO--setup during the observations}
\label{ovro_obs}
%the * means that it will be printed in 2collumn format
\begin{center}
\small
\begin{tabular}{lccc}
\tableline \tableline
OVRO configuration & C & L & H \nl
baselines & $7-22\,\mbox{k}\lambda$ & $6 - 44\,\mbox{k}\lambda$ & $13 - 92\,\mbox{k}\lambda$\nl
& $ (20-60 \,\mbox{m})$ &  $ (15-115 \,\mbox{m})$ & $ (35-241
\,\mbox{m})$\nl
date of observation & 1999 Sep.\ 22 & 1999 Nov.\ 10 & 1999 Dec.\ 16  \nl
                    & 1999 Sep.\ 24 & 1999 Nov.\ 13 & 1999 Dec.\ 17  \nl 
total time on source & 800 min & 900 min & 950 min \nl
\tableline
total bandwidths &  & 124 MHz, 31 MHz &  \nl
No. of channels &  & 62, 62 &  \nl
velocity resolution &  & 5\,\kms, 1.3\,\kms &  \nl
central velocity &  & 7 \kms &  \nl
angular resolution\tablenotemark{a} & & $ 2.4\times 1.9$ arcsec\nl
linear resolution\tablenotemark{a} &  & $37 \times 29$ pc\nl
rms noise\tablenotemark{a}\tablenotemark{,b} & & 22 mJy\,beam$^{-1}$ (0.44 K)\nl
\tableline \tableline
\end{tabular}
\tablenotetext{a}{based on {\sc robust} data cube; see text}
\tablenotetext{b}{5 \kms\ channel}
\end{center}
\end{table}

\subsection{HI--observations}

NGC~3077 was observed with the NRAO\footnote{The National Radio
Astronomy Observatory (NRAO) is operated by Associated Universities,
Inc., under cooperative agreement with the National Science
Foundation.} Very Large Array (VLA) in B--, C-- and
D--configuration. Part of the D--array observations were affected by
solar interference, which was removed by deleting {\it uv--}spacings
shorter than typically 0.4 k$\lambda$.  In total, 17 hours were spent
on source which were divided into 4 hours for D--array, 2.5 hours in
C--array and 10.5 hours in B--array (see Table~\ref{vla_obs} for a
detailed description of the VLA observations). The flux calibration
was determined by observing 1328+307 (3C286) for approximately 20
minutes during each observing run, assuming a flux density of 14.73 Jy
according to the Baars \ea (1977) scale. This calibrator was also used
to derive the complex bandpass corrections. The nearby calibrators
1031+567 and 0945+664 were used as secondary amplitude and phase
calibrators and their fluxes were determined to be 1.80 Jy and 2.22
Jy, respectivey. Since the systemic velocity of NGC~3077 ($\sim 5$
\kms) overlaps in velocity with Galactic \ion{H}{1} emission, each calibrator observation was observed at velocities shifted by +300 \kms\ and --300
\kms\ to avoid Galactic contamination. In the course of
the calibration these observations were then averaged to give
interpolated amplitude and phase corrections. NGC\,3077 was observed
using a 1.56 MHz bandwidth centered at a heliocentric velocity of 38
\kms. This band was divided into 128 channels resulting in a velocity
resolution of 2.58 \kms\ after online Hanning smoothing. In editing
and analysing the data we followed the same procedure as outlined in
the previous section (including the production of a {\sc master}
cube).  Obtaining fluxes for extended emission in multi-array
observations is non--trivial -- here we followed the procedure
described in Walter \& Brinks (1999).

We also calculated a radio continuum image of NGC\,3077 at 21\,cm by
using the line--free channels of the VLA \ion{H}{1}\ observations;
here we boosted the resolution to 5$''$ ($\sim 77.5$\,pc) by employing
uniform weighting (rms: 0.4 mJy\,beam$^{-1}$).

\begin{table}
\caption{Setup of the VLA during the observations}
\label{vla_obs}
%the * means that it will be printed in 2collumn format
\begin{center}
\small
\begin{tabular}{lccc}
\tableline \tableline
VLA configuration & B & C & D \nl
baselines & $1-54\,\mbox{k}\lambda$ & $0.34 - 16\,\mbox{k}\lambda$ & $0.166 - 4.9\,\mbox{k}\lambda$\nl
& $ (0.21 - 11.4 \,\mbox{km})$ &  $ (0.073 - 3.4 \,\mbox{km})$ & $ (0.035 - 1.03\,\mbox{km})$\nl
date of observation & 1992 Jan.\ 18 & 1992 Apr.\ 20 & 1991 Mar.\ 03  \nl
                    & 1992 Jan.\ 19 &               & 1992 Sep.\ 03  \nl 
total time on source & 647 min & 133 min & 245 min \nl
\tableline
total bandwidth &  & 1.56 MHz &  \nl
No. of channels &  & 128 &  \nl
velocity resolution &  & 2.58 \kms &  \nl
central velocity &  & 14 \kms &  \nl
angular resolution\tablenotemark{a} & & $13.0 \times 12.7$ arcsec\nl
linear resolution\tablenotemark{a} &  & $200 \times 200$ pc\nl
rms noise\tablenotemark{a,b} & & 1.0 mJy\,beam$^{-1}$ (3.7~K)\nl
\tableline \tableline
\end{tabular}
\tablenotetext{a}{based on {\sc natural} data cube}
\tablenotetext{b}{2.58\,\kms channel}
\end{center}
\end{table}

\subsection{Optical and NIR Observations}

Narrowband images were obtained January 2-10, 2000 at the Kitt Peak
National Observatory 2.1m telescope at $\lambda 6487$ (67\AA FWHM) and
$6571\AA$ (FWHM 84\AA).  The CCD frames were corrected for zero level
bias offsets, pixel-to-pixel bias variations, and pixel-to-pixel
sensitivity fluctuations.  After constructing cosmic ray masks for
each frame, and registering the frames, the images in each bandpass
were co-added to obtain on-band and off-band images.  These images
were flux calibrated using observations of spectrophotometric
standards (Massey \etal\ 1988).  The off-band continuum was scaled
iteratively to match the stellar continuum in the on-band image and
then subtracted from it.  The net emission-line image contains both
[NII] $\lambda\lambda 6548, 6583$ and H$\alpha$ emission. The
spectroscopic observations discussed here are described in Martin
(1998).

We retreived NICMOS broad and narrow band images from the HST archive
(PI: Sparks, H: dataset N4K40JP3Q, P$\alpha$: N4K40JP2Q, both images
were obtained on June 16, 1998). In calibrating the images, we
followed the approach given in B\"oker et al. (1999). The continuum
was subtracted from the Pa$\alpha$ images and the flux in the
Pa$\alpha$ line was derived using:$$F_{\rm line} = 1.054 \times FWHM
\times PHOTFLAM \times CR$$ where $FWHM=191\AA$,
$PHOTFLAM=4.685\times10^{-18}$\,erg\,cm$^{-2}$\,\AA$^{-1}$\,DN (the
latest value given in the NICMOS handbook) and CR is the countrate
(DN\,s$^{-1}$).  We have also obtained near infrared broad--band J, H
and K images from the 2MASS project.

\section{Results}

\subsection{The distribution of Neutral Hydrogen}

The distribution of neutral hydrogen at high angular resolution
(moment 0 map, corrected for primary beam attenuation) is presented in
Fig~1, right.  Most of the \ion{H}{1} in the region is situated in the
prominent eastern tidal arm, little \ion{H}{1}\ is associated with the
optical counterpart of NGC\,3077 (contours).

To obtain accurate \ion{H}{1} masses for NGC\,3077 we defined 3
different apertures as indicated in Fig.~1. The largest aperture (I)
accounts for most of the \ion{H}{1} in the field of view. Aperture II
only includes \ion{H}{1} which is roughly associated with the optical
body of NGC\,3077 --- the smallest aperture (III) only accounts for
the central nucleus. The total fluxes (and masses, adopting a distance
to NGC\,3077 of 3.2\,Mpc) for the apertures are: (I): 192.3 Jy \kms\
($4.8 \times 10^8$ \msun), (II): 42.13 Jy \kms\ ($1.0 \times 10^8$
\msun) and (III): 11.30 \kms\ ($2.8 \times 10^7$ \msun). The
\ion{H}{1}\ content associated with the optical body of NGC3077 is therefore
somewhere between $3-10 \times 10^7$ \msun\ (depending on the aperture
chosen).

As mentioned in the Introduction, the massive \ion{H}{1} complex east
of NGC\,3077 has presumably been stripped off the outskirts of
NGC\,3077 during the recent interaction with M\,81. Evidence for this
comes from numerical simulations (e.g., Yun et al. 1993) which show
that the material in the tidal arm east of NGC\,3077 originally
belonged to NGC\,3077 itself. This scenario is supported by the
high--resolution \ion{H}{1} observations: Figure~2 shows the
\ion{H}{1} velocity field of NGC\,3077 and the region around it. The
mean velocity of the tidal feature is $v=15$\,\kms\ and close to the
one of NGC\,3077 ($v_{\rm sys}=5$\,\kms). Note that the systemic
velocity of M\,81 is some $v_{\rm sys}=50$\,\kms\ with velocities
reaching $v=-250$\,\kms\ in the spiral arms pointing towards NGC\,3077
(Westpfahl \& Adler 1996). We also find no evidence for \ion{H}{1}\
mass accretion and/or infall onto the centre of NGC\,3077 in our
\ion{H}{1} data. All this renders the possibility that
the material originally belonged to M\,81 unlikely. An additional (but
somewhat weaker) argument comes from the fact that the \ion{H}{1}\
mass to blue light ratio in centre of NGC\,3077 is only $\sim 0.05$
M$_{\odot}$/L$_{\rm B\odot}$ (see Tab.~1). If the \ion{H}{1} in the
tidal arm previously belonged to NGC\,3077 the \ion{H}{1}\ mass to
blue light ratio is more like $\sim 0.5$M$_{\odot}$/L$_{\rm B\odot}$,
more typical for dwarf irregular galaxies (this argument obviously
only holds if NGC\,3077 was not a dwarf elliptical before the
interaction). We conclude that the tidal gas around NGC\,3077
originated from NGC\,3077 itself. We note that since the total mass of
the tidal system is similar in mass to NGC\,3077 itself (Heithausen \&
Walter, 2000), a complete detachment of this system (i.e., the
creation of a true tidal dwarf system) is likely (but see Meier,
Turner \& Beck 2001).

The lines in Fig.~2 indicate the orientation of 3 position velocity
(pv) diagrams through the \ion{H}{1}\ distribution around NGC\,3077 as
shown in Fig 3.  Multiple velocity components are visible where the
\ion{H}{1} emission in the tidal feature is strongest. The cuts along
the tidal arms towards M\,81 and M\,82 show a rather smooth velocity
gradient away from the the \ion{H}{1} maximum in the tidal feature.
The first pv diagram (Fig.~3, I) cuts through the center of NGC\,3077 and
the prominent tidal feature. The bright emission at offset 2.3$'$ is
NGC\,3077 itself. Multiple velocity components are visible in the
tidal feature (around offset -2$'$). The second (II) and third (III)
pv diagrams are oriented along the prominent tidal arms which roughly
point towards M\,82 (north, cut II) and M\,81 (west, cut III). They
show a rather smooth velocity gradient away from the \ion{H}{1}
maximum in the tidal feature.

\subsection{The Distribution of Molecular Gas \label{co}}

\subsubsection{Global properties}

The CO(1--0) channel maps as obtained with OVRO are shown in Fig.~4
(based on the natural weighted data at 1.3\,\kms\
resolution). Clearly, the CO emission in NGC\,3077 originates from
many different sub--structures. These sub--clumps are even better
visible in position velocity diagrams as the one presented in Fig.~5
(at high spatial and velocity resolution -- the orientation of this
cut is indicated in Fig.~6, left). The numbers in this plot refer to
the individual subclumps discussed in Sec.~3.2.2. The CO intensity
distribution (moment 0) is presented in Fig.~6 (left: natural
weighting, emphazising diffuse emission, right: robust weighting,
focussing on the small scale structure). Globally, we detected 4
molecular regions which are labeled R1 -- R4 in Fig.~6.

\subsubsection{Clump decomposition and molecular cloud masses}

R1 -- R3 have a complex velocity structure and consist of
multi--velocity components. To analyze the substructure of these
molecular complexes we decomposed the robust weighted ($\sim2''$, high
velocity (1.3 \kms) resolution data cube with a gaussian clump fitting
code (Stutzki \& G\"usten 1990) as well as by eye. The clump fitting
code (`Gaussclump') determines the maximum in the data cube and
subtracts a 3D gaussian fit to the data (RA, DEC, velocity) from the
cube.  The procedure works iteratively on the residual until the
specified RMS level (5 sigma in this case) is reached. In this
procedure the spatial and velocity resolution is taken into account
(for further detail see Stutzki \& G\"usten 1990). Each clump
determined by `Gaussclump' was checked by eye using pv-cuts and by
inspecting spectra at individual positions. In case of doubt we reran
Gaussclump with different parameters for contrast, minimum spatial and
velocity structures and other parameters controlling the gaussian
fit. The parameters for our final clump decomposition of the data set
are given in Tab.~4. Parameters regarding the spatial and velocity
information as well as the flux (column 2 to 7) refer to the fit
results. In total we identified 16 CO complexes within NGC\,3077
(Tab.~4). To emphasize the complex substructure we show a pv-diagram
(along the line shown in Fig.~6, left) in Fig.~5. Nine out of the 16
components are at least partly visible in this diagram and are labeled
by their respective number. The size of most of the clumps is
comparable to our beam implying that most of the CO emission is not
resolved at a linear resolution of $\sim30$\,pc. The radius and the
velocity width shown in Tab.~4 are deconvolved for the beam--size as
well as the velocity--width. For measured values less than 1.1 times
the resolution, the estimate is listed as an upper limit.  The
approximate errors for the measured quantities are given in the
footnote of Tab.~4.  We derived cloud masses assuming that the clouds
are virialized using $M_{\rm vir}=250\times v_{\rm FWHM}[{\rm
\kms}]^2 \times R[{\rm pc}]$ (Rohlfs \& Wilson, 1996; see Tab.~4 column 8). 
We also derived molecular masses assuming a Galactic conversion factor
($X_{CO} = 2.3\times10^{20}\,{\rm cm}^2\,{\rm K}^{-1}\,{\rm
km}^{-1}\,{\rm s}$, Strong \etal\ 1988) using $M=1.23\times10^4\times
(3.2)^2\times S_{CO}\,M_{\odot}$ (column 9).  Complexes 1 to 5 are the
components associated with region R1, complexes 6 to 11 belong to R2,
complexes 12 to 15 belong to R3 and complex 16 is the diffuse CO
component in the north--west of NGC\,3077 (R4).  The virial masses for
clouds in R1--R3 are significantly lower than the masses derived from
converting CO luminosities to \hh\, column densities (using the
Galactic conversion factor). In column 10 we present the derived
conversion factor X$_{\rm CO}$ based on the assumption that the
complexes are indeed virialized. On average, we get values below
$1.0\times 10^{20}\,$cm$^{-2}\,($K\,km\,s$^{-1})^{-1}$ (see the
discussion in Sec.~4). We note that Meier, Turner \& Beck (2001)
derived a Galactic conversion factor for NGC\,3077 (see their
paper for details).

In the following we will adopt a conversion factor of $1\times 10^{20}
{\rm cm}^{-2} ({\rm K} \kms)^{-1}$ for NGC\,3077 (roughly half of the
Galactic value), primarily for convenience. Using this conversion
factor, the total \hh\, mass of NGC\,3077 is $1.6\times10^{6}\, \msol$
(this is counting the complexes discussed above as well as diffuse
emission that has not been catalogued).  Becker et al.\,(1989)
observed the center of NGC\,3077 with the IRAM 30\,m telescope in the
CO($1\to0$) and CO($2\to1$) transition at resolutions of 21$''$ and
13$''$, respectively. They only present the CO($2\to1$) data in their
paper and conclude that a giant molecular complex with a velocity
width of 30\,\kms\ and a diameter of 320\,pc (FWHM) is located near
the center. Using a conversion factor of $4\times 10^{20}\, {\rm
cm}^{-2} ({\rm K} \kms)^{-1}$ they derive a total molecular mass of
$1\times10^7$\,\msun\ and a virial mass of $3\times10^7$\,\msun.
Obviously, our higher resolution observations show that their complex
breaks up into many smaller complexes. The linewidth used by Becker
\etal to derive the virial mass therefore measured the velocity
dispersion of the CO clumps rather than the intrinsic linewidth of a
gravitational bound complex -- their virial mass is therefore too
high. Nevertheless our mass estimate based on the measured CO
intensity of $1.6\times10^{6}\,\msol$ is about a factor of 1.5 less
than the \hh\, mass derived by Becker \etal (taking the different
conversion factors into accout). This indicates that we may miss some
extended emission in our interferometer data. \\

{\small
\begin{table}
\caption{Properties of the CO clumps in NGC\,3077}
\label{co-clumps}
\hspace*{-2cm}
\begin{tabular}{lllccccccc}
\tableline \tableline

&RA (2000) & DEC (2000) & v$_{lsr}$& dv$_{\rm FWHM}^{a}$& r$^{a}$ & $S_\nu$  & M$_{vir}$ & M$_{H_{2}}^{b}$ & X$_{\rm co}^{c}$\nl
&&&[\kms]&[\kms]&[pc]& [Jy km\,s$^{-1}$] &[$10^4 $ \msol] &[$10^4 $ \msol] 
&[$10^{20}\,{\rm cm}^2\,{\rm K}^{-1}\,{\rm km}^{-1}\,{\rm s}$] \nl
1&10:03:18.95 & 68:43:57.37& 24.5 & 4.5 & $<17$    & 1.15 & $<8.4$ & 14.5& $<1.3$ \nl
2&10:03:18.67 & 68:43:55.38& 21.9 & 1.2 & $<17$    & 0.42 & $<0.6$ & 5.2 & $<0.3$ \nl
3&10:03:19.32 & 68:44:00.36& 19.2 & 2.1 & $<17$    & 0.62 & $<1.8$ & 7.8 & $<0.5$ \nl
4&10:03:18.86 & 68:43:56.88& 15.5 & 6.2 & 24       & 3.36 & 22.1   & 42.4& 1.2     \nl
5&10:03:19.14 & 68:44:00.37& 11.6 & 3.4 & $<17$    & 1.07 & $<4.7$ & 13.5 & $<0.8$ \nl
6&10:03:18.77 & 68:43:56.38&  6.4 & 3.1 & $<17$    & 0.92 & $<4.0$ & 11.6 & $<0.8$ \nl
7&10:03:18.95 & 68:44:01.37& -2.8 & 1.9 & 18       & 0.62 & 1.5    & 7.8 & 0.4     \nl
8&10:03:18.78 & 68:44:00.38& -8.0 & 2.5 & $<17$    & 0.56 & $<2.5$ & 7.0 & $<0.8$  \nl
9&10:03:18.86& 68:43:59.88&-13.0 & 1.6  & $<17$    & 0.55 & $<1.0$ & 7.0 & $<0.3$  \nl

%10&10:03:18.86& 68:43:59.88&-15.7 & 3.6 & $<17$    & 1.05 & $<5.4$ & 13.2 & $<0.9$\nl

\tableline
10&10:03:20.33 & 68:44:01.82 &  3.8 & 1.4 & $<17$  & 0.34 & $<0.8$ & 4.3 & $<0.4$ \nl
11&10:03:20.24 & 68:44:04.33 &  3.7 & 2.2 & $<17$  & 0.55 & $<2.0$ & 6.9 & $<0.7$ \nl
12&10:03:20.24 & 68:44:03.83 &  2.5 & 1.9 & $<17$  & 0.60 & $<1.5$ & 7.6 & $<0.5$ \nl
13&10:03:19.96 & 68:44:02.34 &  2.5 & 2.1 & $<17$  & 0.62 & $<1.8$ & 7.8 & $<0.5$ \nl
14&10:03:20.15 & 68:44:03.83 & -2.8 & 2.0 & $<17$  & 0.54 & $<1.6$ & 6.8 & $<0.6$ \nl
15&10:03:20.89 & 68:44:04.30 & -4.0 & 1.8 & $<17$  & 0.51 & $<1.4$ & 6.4 & $<0.5$ \nl
\tableline
16&10:03:14.10 & 68:44:09.05 & -21.3 & 7.9 & 103        & 18.4 & 153  & 232 &  1.5\nl
\tableline \tableline
\end{tabular}
\\
a) deconvolved\\ 
b) This assumes a 'standard' CO-to-H$_2$ conversion
factor of $2.3\times10^{20}$\,cm$^{-2}$\,(K\,km\,s$^{-1}$)$^{-1}$,
Strong et al.\ (1988) including a correction for helium. If the
recalibrated conversion factor of $X = 1.6
\times10^{20}$\,cm$^{-2}$\,(K\,km\,s$^{-1}$)$^{-1}$, Hunter et al.\
(1997) is used, all CO-based masses have to be reduced by $\sim 30$\%. \\
c) assuming M$_{H_{2}}$ = M$_{vir}$\\
Note on errors: the measured quantities have the following errors: 
$\sigma$(v$_{lsr}$,dv)$\sim$0.6\,\kms, $\sigma$(r)$\sim$
10\,pc, $\sigma$(S$_{\nu}$) $\approx 20\%$\\
\end{table}
}

\subsection{Comparison between the atomic and molecular gas phase}

The CO and \ion{H}{1} distribution are compared in Fig.~7 (left). The
greyscale and the thin countours represent the \ion{H}{1} surface
brightness.  The thick contours represent the CO distribution as shown
in Fig.~4 (right). Note the apparent asymmetry between the
\ion{H}{1} and the CO distribtion: the \ion{H}{1} maxima is located
east of the CO maximum. Most of the CO emission is located in an
\ion{H}{1} depression -- this suggests that atomic material in this
region has largely turned molecular.  The total \ion{H}{1} mass in the
area shown is $\sim8\times10^6$\,\msun. 

We created a total gas column density map by combining the \ion{H}{1}
and CO moment maps (Fig.~10, right). The CO data were first convolved
to the resolution of the \ion{H}{1} map ($13.1''\times12.7''$) and the
H$_2$ column density derived from employing a conversion factor
(Sec.~3.2.2) was multiplied by two to get the proton density. The
molecular gas has a clear inpact on the total gas column density map
(reaching maximum surface densities of 5.2$\times 10^{21}$
cm$^{-2}$). The total gas mass (atomic and molecular) in this plot is
$\sim9\times10^6$\,\msun.

The thick contours represent the orientation of the J--band image
(2MASS) of NGC\,3077\footnote{note that the registration of the
2.2$\mu$m image presented in Price \& Gullixson (1989) is off by
$\sim10''$}. The J--band image mostly shows the distribution of the
old stellar population of NGC\,3077 (see Sec.~3.3). It is evident that
the apparent asymmetry of the gas discussed above is weaker if one
considers the total gas column density.  Nevertheless the gas
distribution in the center of NGC\,3077 is clearly displaced towards
the south--east compared to the stellar distribution (tracing the
potential of the galaxy). This implies that not only the gas in the
outer parts of NGC\,3077 has been affected by the tidal forces within
the triplet but also the gas in the center.  A similar displacement of
the gaseous component with respect to the nucleus has been found in
the center of M\,82 (Wei\ss\ \etal\ 2001).

\subsection{Distribution of Stars and Dust}

The optical appearance of NGC\,3077 is heavily influenced by dust
absorption near the nucleus (see the study by Price \& Gullixson
1989). In Fig.~8 we present a B--band image of NGC\,3077. The contours
again represent the J--band image of NGC\,3077 as obtained from the
2MASS survey.  A NICMOS H--band image obtained on board HST of the
same region is presented on the right side of this figure. The near
infrared image mainly shows the distribution of the older stellar
population in NGC\,3077. 

Fig.~9 (left) shows the same broad band image as Fig.~8. The contours
in the left panel of Fig.~9 are the OVRO CO contours as presented in
Fig.~6. The dust feature in the south coincides very well with the
regions R1 and R2 where we detected molecular gas. This finding
implies that CO is a good tracer for dust in NGC\,3077. Price \&
Gullixson (1989) derive an \hi\ mass of $10^5$\,\msun\ for the
absorbing cloud.  Our CO--data implies a total virial mass of
$<5\times 10^5$\,\msun\ (only summing up the complexes C1 to C11). The
difference might indicate that not all clumps (1-11) are associated
with the foreground dust extinction. This view is supported by the
fact that this particular region shows a large CO velocity gradient,
suggesting a large distance between individual clouds along the line
of sight (see Sec.~3.5).

The optical images of NGC\,3077 also reveal the presence of radial
absorption features (indications can be already seen in Fig.~9,
left). The dust lanes look similar to the spokes of a cartwheel; they
are aligned radially in the north--west, however they look more
`curved' in the south--east (see the discussion in Sec.~4). Some of
these dust lanes are associated with weak H$\alpha$ emission as far
out as 500\,pc from the center. To reproduce this situation we present
5 intensity profile cuts parallel to the right ascension in B--band as
well as in H$\alpha$ emission. The intensity cuts are indicated as 5
lines in Fig.~11 and are plotted in Fig.~9 (right). The lines showing
the absorption represent the B--band emission along one cut; the
H$\alpha$ intensity profiles show emission.

\subsection{True location of the starburst: H$\alpha$ and Pa$\alpha$ observations}

Major parts of the ongoing star formation in NGC\,3077 as traced by
H$\alpha$ emission is hidden by the central dust cloud (Sec.~3.3). To
reveal the true location of the starburst we analyzed Pa$\alpha$
imaging (NICMOS, HST).  The Pa$\alpha$ image is presented in Fig.~10
(right) -- our ground--based H$\alpha$ image of the same region is
shown on the left side of the figure. As discussed in Martin (1998)
the central starburst region is surrounded by expanding H$\alpha$
shells which are breaking out from the center (these shells are also
visible in Pa$\alpha$).  The brightest \ha\, emission is located just
north of the molecular complexes R1 and R2.  Thronson
\etal\,({THR89}1998) suggested that this morphology might be
indicative for a confinement of the ionized gas to the south and east
where the molecular complexes and the \hi\, column density peak are
located. However, from the Pa$\alpha$ imaging it is obvious that major
parts of the starburst are hidden by the dust feature discussed in
Sec.~3.3.  The contours in the right panel of this Fig.~10 again
represent the CO emission from our high--resolution OVRO data (see
Fig.~13 for a color composite).

The Pa$\alpha$ emission peaks right in between the regions R1 and R2
where only a minor fraction of \ha\ is detected. This implies that due
to absorption the \ha\ emission in NGC\,3077 only partly traces the
ongoing star formation.  This view is supported by our observations in
the radio continuum at 21\,cm which are shown as contours in the left
panel of Fig.~10.  Although our resolution of 5$''$ ($\sim80\,$pc) is
too poor to resolve the emission, the radio continuum emission clearly
shows that the most active starforming region is associated with the
molecular complexes R1 and R2. A comparison between the CO kinematics
and our \ha\ slit spectroscopy (Fig.~12, the orientation of the slit
is indicated in Fig.~10, left) in this particular region reveals
another interesting aspect: the center velocity of the \ha\ emission
is between the velocities of the molecular complexes R1 and R2. It is
suggestive that this region of strong star formation has disrupted its
parental molecular cloud which was located between R1 and R2 . Note
that even the faint diffuse CO complex R4 in the north--west
(Sec.~3.2.1) coincides with H$\alpha$ emission at the same velocity.

A quantitative analysis of the narrow band images shown in Fig.~10
yields a total Pa$\alpha$ flux of NGC\,3077 of
F(Pa$\alpha$)=$7.8\times10^{-13}$\,erg\,cm$^{-2}$\,s$^{-1}$
(consistent with the value derived by B\"oker \etal\ 1999). The total
H$\alpha$ flux is F(H$\alpha$)=$4.4\times10^{-12}$
\,erg\,cm$^{-2}$\,s$^{-1}$ (L(H$\alpha$)=$5.33\times 10^{39}$\,erg\,s$^{-1}$, 
using $L=4\,\pi\,D^2F(H\alpha)$). 

The Pa$\alpha$/H$\alpha$ intensity ratio in the outer parts of
NGC\,3077 (where the absorption of H$\alpha$ is negligible) is about
0.17. This is close to the ratio of the emission coefficients for
$T=10^4$\,K and Case~B recombination of $j($Pa$\alpha)/j($Ha$\alpha) =
0.12$ (Osterbrok 1989, Table 4.4).  The highest absorption in the
galaxy is found close to the Pa$\alpha$ peak (intensity ratio
(Pa$\alpha$/Ha$\alpha$) = 1.30).  This corresponds to a local
extinction for this particular region of about an order of magnitude
($\sim$2.5\,mag).

If there was no extinction we would expect an H$\alpha$ flux of
5.2$\times10^{-12}$\,erg\,cm$^{-2}$\,s$^{-1}$ based on the observed
Pa$\alpha$ flux (L(H$\alpha$)=$6.30\times 10^{39}$\,erg\,s$^{-1}$).
We used
SFR(\msun\,yr$^{-1}$)=L(H$\alpha$)/($1.26\times10^{41}$)\,erg\,s$^{-1}$
(Kennicutt \etal\ 1994) to derive a star formation rate (SFR) based on
the extinction corrected H$\alpha$ emission of
0.05\,\msun\,yr$^{-1}$. This is some 25\% larger than the value
obtained from the H$\alpha$ emission alone and is a measure for the
total extinction within NGC\,3077. Note that this is a factor of a few
less than the value derived by Meier, Turner \& Beck (2001) who used
2.6\,mm radio continuum emission to derive the SFR in NGC\,3077.

\subsection{Outflow of Ionized Gas vs. \ion{H}{1}\ morphology}

The outflow in NGC\,3077 (as traced by H$\alpha$ and Pa$\alpha$
emission) reaches galactocentric distances of up to 1.5\,kpc. Given
the intrinsic small size of NGC\,3077 this is remarkable (for a
detailed discussion see Martin 1997).  At larger scale, an
anti--correlation between diffuse \ion{H}{1} and the H$\alpha$
outflow is evident: In Fig.~11 we plot the
\ion{H}{1}\ in blue and the H$\alpha$ in red (for comparison, the
optical broad band image is shown in green as well). At large
galactocentric radii, the \ion{H}{1}\ is predominantly situated $
around$ the expanding shells. Typical \ion{H}{1} surface densities at
these galactocentric radii are $4-9\times10^{20}$\,cm$^{-2}$.  One
explanation for these feature may be that the pressure of the interior
of the H$\alpha$ shells is high enough to push the \ion{H}{1} out to
larger radii. 

\section{Discussion and Summary}

The wealth of data presented here give exciting new insights on the
interplay between the atomic gas, molecular gas and ongoing star
formation in NGC\,3077. The total (atomic and molecular) gas content
in the centre of NGC\,3077 is clearly displaced from the stellar
component of NGC\,3077 -- this implies that not only the gas at large
galactocentric radii is affected by the interaction within the
triplet. The \ion{H}{1}\ mass associated with NGC\,3077 is
$3-10\times10^7$\msun (depending on the aperture chosen, Sec.~3.1). It
was suggested in numerical studies of the triplet that the tidal
material around NGC\,3077 originally belonged to NGC\,3077 itself; the
high--resolution data of the \ion{H}{1} morphology and kinematics show
observational evidence for this scenario.  

Based on the OVRO interferometer data we estimate a molecular mass of
$1.6\times10^6$ \msol for NGC\,3077 (Sec.~3.1.2). In total, 16
complexes have been detected with our 2$''$ ($\sim 30$\,pc) beam.  All
of the complexes have virial masses which are lower than the masses
calculated from the CO luminosity (assuming a Galactic conversion
factor). This implies that the conversion factor for NGC\,3077 is
lower than the Galactic value.  At first glance, this seems surprising
since the conversion factor is usually believed to be higher in dwarf
galaxies as compared to spirals such as our own Galaxy (Arimoto \etal\
1996, Wilson 1995). However, the metallicity of NGC\,3077 is high (O/H
around solar) for a dwarf (Martin 1998), so this objection may not
apply. The higher CO emissivity and lower conversion factor can also
be understood by an increased kinetic temperature and cosmic--ray
heating near the starburst powering the outflow in NGC\,3077. E.g., in
the case of M\,82, Wei\ss\ et al. (2001) have shown that X$_{\rm CO}$
may be as low as
$3\times10^{19}\,$cm$^{-2}\,($K\,km\,s$^{-1})^{-1}$. The prominent
star forming regions as well as the outflow visible in H$\alpha$
and Pa$\alpha$, as well as the radio continuum source near the
starburst already suggest that the state of the molecular gas in
NGC\,3077 may be affected by the starburst (similar to the case in
M\,82). Indeed, as already pointed out by Becker \etal\ (1989), a
CO(2-1)/CO(1-0) line intensity ratio of 0.82 (on a T$^*_{A}$ scale)
indicates that a substantial fraction of the gas is heated to kinetic
temperatures $>\,10$\,K. In a more recent study Meier \etal\ 2001,
using observations in the CO(3-2) transition, estimated a kinetic gas
temperature in NGC\,3077 of T$_{kin}\approx 30$\,K using LVG
calculations. We note that the CO(2-1)/CO(1-0) line intensity ratio of
0.82 as reported by Becker \etal\ (1989) refers to antenna
temperatures. Taking antenna efficiencies into account the
CO(2-1)/CO(1-0) line intensity ratio is about 1.3 (T$_{mb}$ scale,
efficiencies as in Guelin \etal\ 1995) indicative for even larger
temperatures. Using the ratio above and CO(3-2)/CO(1-0)\,=\,0.7 (Meier
\etal\ 2001; corrected for efficiencies at 115\,GHz) we find T$_{kin}>
35$\,K using LVG models. This supports the idea that the molecular gas
in NGC\,3077 is heated by the starburst resulting in a lower
conversion factor (cf. Weiss et al.\ 2001 in the case of M\,82).

We find a striking correspondence between the CO distribution and the
dust cloud seen in the optical. From the Pa$\alpha$ and H$\alpha$
imaging we estimate the extinction in the central dust complex to be
an order of magnitude ($\sim 2.5$\,mag). Surprisingly, CO emission is
also coincident in position and velocity with a faint star forming
region located 500\,pc west of NGC\,3077, far off the central star
formation.

A comparison of the ionized outflow with the \ion{H}{1}\ shows that
the distribution of \ion{H}{1}\ around NGC\,3077's center is affected
by the outflow.  A puzzle is the presence of the `radial'
dust--fingers which reach lengths up to $\sim 1$\,kpc. It may be that these
dust lanes originally belonged to rudimentary spiral arms which have
been ripped apart by the interaction.  The dust lanes may have also
been blown out of the NGC\,3077's center by the violent SF (e.g., by
radiation pressure).  Somewhat counter--intuitively, the most
prominent lanes are associated with H$\alpha$ emission. One way to
interpret the apparent correlation of dust with H$\alpha$ is that the
surface of the dust filaments is exposed to the strong radiation field
of the central starforming region which causes the weak \ha\, emission
on the rim of the filament.

Correcting for extinction, we derive a star formation rate (SFR) of
0.05\,\msun\,yr$^{-1}$ for NGC\,3077. Although this value does not
look dramatic one has to keep in mind that the star formation in
NGC\,3077 is very concentrated towards the center. The effective area
where massive star formation is found is $\sim2\times 10^4$\,pc$^2$
resulting in a SFR per surface area of
$2\times10^{-6}$\msun\,yr$^{-1}$pc$^{-2}$, much higher than what is
usually found in spiral galaxies (e.g. Martin \& Kennicutt 2001).

We will now compare the neutral hydrogen, molecular gas and starburst
properties of NGC\,3077 to the corresponding parameters of the
prototypical starburst galaxy M\,82 (situated at a projected distance
of only $\sim 70$\,kpc).  About 10\% of the total gas mass (including
the tidal arm) of NGC\,3077 is associated with the optical galaxy; the
corresponding fraction for M\,82 is $\>$70\% (Yun et al.\ 1993b). This
suggests that NGC\,3077 lost most of its gas during the interaction
whereas M\,82 was able to hold it. The central starbursting region of
M\,82 still has a reservoir of about $2\times10^8\msol$ of neutral
hydrogen (more than $\sim$4 times the corresponding value of
NGC\,3077); the molecular gas content of M\,82 is
$2.2\times10^8$\,\msun ($\sim140$ times higher than in NGC\,3077,
Wei\ss\ et al.\ 2001, D$_{\rm M82}$=3.2\,Mpc). The ratio of molecular
to atomic gas in M\,82's center is therefore of order unity,
significantly higher than the same ratio in NGC\,3077 ($\sim0.05$).
This implies that the efficiency to form molecular out of atomic gas
is much higher in M\,82 as compared to NGC\,3077.  Interestingly the
star formation efficiencies (SFE) are similar in NGC\,3077 and M\,82:
our \ha\ and CO observations yield a SFE for NGC\,3077 of
L$_{\ha}$/M(\hh) $\approx 1\,\lsol\,\msol^{-1}$
(\lsol=$3.82\times10^{33}$\,erg\,s$^{-1}$. The corresponding value for
M\,82 is about $\approx 0.3\,\lsol\,\msol^{-1}$ (adopting L$_{\ha} =
2.3\times10^{41}$\,erg\,s$^{-1}$; Heckman \etal\ 1990, the true
H$\alpha$ flux in the centre of M\,82 is presumably somewhat higher
due to the edge--on orientation of M\,82).

In that sense NGC\,3077 can be regarded as a scaled--down version of
M\,82: it has a similar star formation efficiency and shows a
prominent outflow. The `only' differences may be that the overall
total masses are different and that NGC\,3077 lost a larger fraction
of its atomic gas during the interaction as M\,82 did.  In any case
the starburst activity in both galaxies was presumably triggered by
the redistribution of atomic and molecular gas in their centers due to
the gravitational interaction within the M\,81 triplet.

\acknowledgments

FW acknowledges NSF grant AST96--13717.  AW acknowledges DFG grant SFB\,494. 
CLM acknowledges support from a Sherman Fairchild Fellowship. 
Research with the Owens Valley Radio Telescope, operated by Caltech, 
is supported by NSF grant
AST96--13717. Support of this work was also provided by a grant from
the K.T. and E.L. Norris Foundation. KPNO/NOAO is operated by the
Association of Universities for Research in Astronomy (AURA),
Inc. under cooperative agreement with the National Science
Foundation. The National Radio Astronomy Observatory (NRAO) is
operated by Associated Universities, Inc., under cooperative agreement
with the National Science Foundation. This research has made use of
the NASA/IPAC Extragalactic Database (NED) which is operated by the
Jet Propulsion Laboratory, Caltech, under contract with the National
Aeronautics and Space Administration (NASA), NASA's Astrophysical Data
System Abstract Service (ADS), and NASA's SkyView.  The 2MASS project
is a collaboration between The University of Massachusetts and the
Infrared Processing and Analysis Center (JPL/Caltech).

\newpage

\begin{figure}
\epsscale{1}
%\plotone{walter.fig1_new.ps}
\figcaption{ {\em Left:}
High--resolution VLA \ion{H}{1} map of NGC\,3077 and its environment.
The contours represent the optical image (shown in greyscale on the
right). The three ellipses define the apertures used for the
\ion{H}{1}\ mass determination. Note that most of the \ion{H}{1}\ is
not associated with the optical body of NGC\,3077 but situated in a
extended tidal arm towards the east.  {\em Right:} Optical broadband
image of the same region as shown on the left. The two boxes (labeled
A, diameter: 2~kpc, and B, diameter: 1~kpc) are the regions which will
be discussed in the next figures.}
\end{figure}

\begin{figure}
\epsscale{0.75}
%\plotone{walter.fig2_new.ps}
\figcaption{Velocity field of \ion{H}{1} around NGC\,3077. Velocities range from --60 \kms (blue) to +40\kms (red). The box indicates the region shown in Fig.~1 (left). The black contours again represent the
optical image. The 3 lines indicate the orientation of the position
velocity diagrams shown in Fig.~3.}
\end{figure}

\begin{figure}
\epsscale{0.4}
%\plotone{walter.fig3_new.ps}
\figcaption{\ion{H}{1} position velocity (pv) diagrams along 3 cuts through the tidal arms around NGC\,3077 (see Fig.~2 for the orientation). The velocity resolution in these diagrams is $2.52$\kms. The
first pv diagram (I) cuts through the center of NGC\,3077 and the
prominent tidal feature. The bright emission at offset 2.3$'$ is
NGC\,3077 itself. Multiple velocity components are visible in the
tidal feature (around offset -2$'$). The second (II) and third (III)
pv diagrams are oriented along the prominent tidal arms which roughly
point towards M\,82 (north, cut II) and M\,81 (west, cut III).
}
\end{figure}

\begin{figure}
\epsscale{0.8}
%\plotone{walter.fig4_new.ps}
\figcaption{Channel maps of the natural weighted OVRO CO observations. Contour levels are 2, 4, 6, 8 $\times \sigma$ (1$\sigma=35\,$mJy\,beam$^{-1}$, 0.29\,K, beamsize: 3.7$\times$3.0$''$, channel seperation:~1.3\kms). }
\end{figure}

\begin{figure}
\epsscale{0.8}
%\plotone{walter.fig5_new.ps}
\figcaption{Position--velocity diagram through the high--resolution OVRO CO cube (as indicated by a line in Fig.~6. Note that the emission is breaking up in many smaller subclumps. The clumps listed in Tab.~4 which are visible in this diagram are labeled by their number (column~1, Tab.~4).}
\end{figure}

\begin{figure}
\epsscale{0.8}
%\plotone{walter.fig6_new.ps}
\figcaption{
Integrated CO--map of NGC\,3077 (the area shown is indicated by box B
in Fig.~1, right). {\em Left:} based on the {\sc natural} data
(resolution $3.7'' \times 3.0''$, $\sim 50$\,pc), the line indicates the orientation of the position velocity cut shown in Fig.~5 {\em Right:} based
in the {\sc robust} data ($2.4'' \times 1.9''$, $\sim 30$\,pc). R1--R4
label the four regions discussed in the text.}
\end{figure}

\begin{figure}
\epsscale{1}
%\plotone{walter.fig7_new.ps}
\figcaption{{\em Left:} Integrated HI--map of the center of NGC\,3077. The thin
contours represent the \ion{H}{1} surface density at 1, 1.5 and
2$\times10^{21}$\,cm$^{-2}$. The thick contours represent the
integrated CO emission (area as indicated by B box in Fig.~1, right).
{\em Right:} Total proton column density map (\ion{H}{1}+CO, convolved
to the beamsize of \ion{H}{1}). Contours are plotted at 1,2,3,4 and
5$\times10^{21}$\,cm$^{-2}$. The thick contours represent NGC\,3077 in
J--band (2MASS survey). Note that the apparent assymetry gas/galaxy is
weaker if one considers the total proton density.
\label{hi-co-map}} 
\end{figure}

\begin{figure}
\epsscale{1.0}
%\plotone{walter.fig8_new.ps}
\figcaption{{\em Left:} B--band image of NGC\,3077 -- the contours represent a J--band image as obtained from the 2MASS catalog (same contours as Fig.~7, right). Note that the two maxima do not coincide which indicates dust absorption south of the nucleus. {\em Right:} H--band image as obtained with NICMOS on board HST (area shown in both panels is indicated by box B in Fig.~1).
\label{co_spec}} 
\end{figure}

\begin{figure}
\epsscale{1.0}
%\plotone{walter.fig9_new.ps}
\figcaption{{\em Left:} B--band image of NGC\,3077. The area shown is indicated by box B in Fig.~1, right. The contours are the high--resolution CO data. The CO emission beautifully coinides with the aborption feature south of the nucleus. {\em Right:} 5 intensity cuts (B--band as well as H$\alpha$) along the lines shown in Fig.~11. The intensity (A.U.) is plotted versus position. Note that the B--band profiles show extinction where the H$\alpha$ shows emission. For presentation purposes we subtracted the
broad underlying H$\alpha$ and B--band emission (by subtracting
baselines). }
\end{figure}

\begin{figure}
\epsscale{1.0}
%\plotone{walter.fig10_new.ps}
\figcaption{{\em Left:} H$\alpha$ image of NGC\,3077 -- the contours represent radio continuum emission at 21\,cm (contours drawn at 2, 3, 4, 5 mJy\,beam$^{-1}$. The line indicates the position velocity cut shown in Fig.~12. {\em Right:} Pa$\alpha$ image of NGC\,3077 as obtained with NICMOS (HST). The contours represent the CO observations. Note that most of the starburst is hidden by the dust absorption feature (as traced by the CO -- see Fig.~8, left). Both sides display the same region as indicated by box B in Fig.~1.  }
\end{figure}

\begin{figure}
\epsscale{0.5}
%\plotone{walter.fig11_new.ps}
\figcaption{Color composite of NGC\,3077: Blue: \ion{H}{1}, Red: H$\alpha$, Green (B--band). The region plotted is indicated by box A in Fig.~1. The 5 yellow lines indicate the positions of the intensity cuts presented in Fig.~8 (right). Note how the H$\alpha$ outflow seems to be confined by the huge \ion{H}{1} halo (see also text).
\label{rgb}} 
\end{figure}

\begin{figure}
\epsscale{0.5}
%\plotone{walter.fig12_new.ps}
\figcaption{Position velocity cuts along the line indicated in
Fig.~9. The greyscale represents the H$\alpha$ velocities -- the
contours are the CO velocities.  \label{pv}}
\end{figure}

\begin{figure}
\epsscale{0.7}
%\plotone{walter.fig13_new.ps}
\figcaption{Color composite of NGC\,3077: Blue: NICMOS H--band, Red: H$\alpha$, Green: CO (OVRO). The region shown is indicated by box B in Fig.~1.}
\end{figure}

\end{document}